| **Title** | **At the Origins of Electroculture: A Retrodictive Modelling of Bertholon's 18th-Century Electrovegetometer in the Pre-Corona Regime** |
|---|---|
| **Authors** | Thierry Dufour[1] |
| **Affiliations** | [1] LPP, Sorbonne Université Univ. Paris 6, CNRS, Ecole Polytech., Univ. Paris-Sud, Observatoire de Paris, Université Paris-Saclay,10 PSL Research University, 4 Place Jussieu, 75252 Paris, France. |
| **Correspondence** | thierry.dufour@sorbonne-universite.fr |
| **Ref.** | Compte-Rendus Mécanique de l'Académie des Sciences. Volume 354 (2026), p. 117-139 |
| **DOI** | https://doi.org/10.5802/crmeca.346 |
| **Summary** | Pierre-Nicolas Bertholon's 18th-century electrovegetometer was conceived to harness "atmospheric electricity" for plant growth, yet its physical capabilities have never been quantified within the context of today's understanding of the Earth's atmospheric electric system. This study addresses the lack of quantitative assessment of such historical "electroculture" device and its plausible influence on the near-canopy electrical environment. It aims to reinterpret Bertholon's apparatus using contemporary atmospheric electrodynamics, asking under which realistic fair-weather and storm-like conditions a purely passive collector-distributor could generate fields and ion fluxes of physical significance. A two-dimensional, quasi-steady ohmic model has been developed in which the atmosphere is a resistive column carrying the global conduction current, the metal structure is a floating conductor supported by leaky wooden insulators and space-charge and corona are excluded so that all results describe pre-onset upper bounds. The simulations show that in fair weather the single upper point and the lower multi-point crown of the electrovegetometer enhance the background field by two to three orders of magnitude, yet only within millimetric-centimetric regions around the tips and with total currents limited to the pA-nA·m$^{-2}$ range. Under storm-like forcing, peak fields at the crown reach $10^5$-$10^6$ V·m$^{-1}$, approaching or exceeding empirical corona-onset thresholds, while remaining largely insensitive to uncertainties in apex angle or collector geometry as long as an elevated mast is present. These results make Bertholon's reports of luminous "aigrettes" physically plausible, but suggest that any fair-weather agronomic impact was subtle and highly localized and that modern "electroculture" claims require careful, coupled electrostatic and biological studies beyond the pre-corona regime. |
| **Keywords** | Electroculture, electrovegetometer, plasma agriculture, Bertholon |

# I. Introduction

## I.1. An interdisciplinary impetus: from the Montpellier colloquium to the present study

The present study was initiated in the aftermath of the French colloquium *« Hommage à Pierre Bertholon, savant et électricien des Lumières »*, held in Montpellier in October 2023 under the patronage of UNESCO. This event gathered historians of science, physicists, atmospheric-electricity specialists and scholars of 18th-century natural philosophy, communities that rarely interact even though their combination holds unsuspected and significant interdisciplinary potential. Discussions with participants during and after the meeting revealed a shared interest in reassessing Bertholon's electrovegetometer using the tools of modern atmospheric electrodynamics, while remaining faithful to its historical context and material constraints. The present article is therefore partly a product of this interdisciplinary encounter, illustrating the scientific value of such colloquia in bridging distinct research cultures and stimulating new lines of inquiry across disciplinary boundaries.

## I.2. Historical framing

To situate this renewed interest within its original scientific context, it is necessary to revisit the historical foundations of Bertholon's work. In 1783, Pierre-Nicolas Bertholon de Saint-Lazare, commonly known as Abbé Bertholon, published *De l'électricité des végétaux*, a treatise in which he asserted that "atmospheric electricity" acts continuously on plants and might be harnessed to improve germination, growth, flowering and fruiting [Bertholon, 1783]. His key practical proposal, the electrovegetometer (*électro-végétomètre)*, was a passive apparatus intended to collect charge from the atmosphere and redistribute it gently over crops through "aigrettes lumineuses": a historical French expression that can be understood today as faint, luminous corona discharges.

Although Bertholon interpreted the operation of his apparatus through the 18th-century notion of an "electric fluid", the physical principle he was appealing to can be expressed clearly in modern electrodynamic terms: sharp conductors immersed in an ambient electric field can act as sources (or sinks) of space charge, modifying local ion densities and weak airflows and thereby altering the microphysical environment near leaves and soil. In Bertholon's descriptions, the electrovegetometer consisted essentially of a vertical, insulated mast that elevated a collecting structure into the atmospheric electric field. A conductive path brought the intercepted charge down to a distribution arm, which terminated in a crown of fine points suspended over crop rows. The points were intended to release charge via non-disruptive discharges, subtly modifying the potential gradient and charge environment around the canopy.

Bertholon described two closely related configurations of this electrovegetometer:

- The Configuration I "single-point rod, articulated arm" (**Figure 1a**), combined a relatively simple collector with a mechanically

versatile diffuser. A tall wooden mast, carefully insulated from the soil by tar, resins and glass elements, carried a single iron point at its head, shielded by a tin funnel to favour upward collection. A suspended chain conducted the intercepted "electric fluid" to a small iron disk where the collected charge was expected to accumulate before being redistributed. From there, a hinged metal arm, horizontally supported on insulated trestles, could be swung over selected rows of crops. At its far end, a crown of fine points could release charge as a diffuse corona just above the canopy. A removable grounding chain allowed the whole device to be neutralized when the air was judged "overcharged".

- The Configuration II "multi-point crown, rotating and extendable arm" (**Figure 1b**) is characterized by a crown of sharp points mounted on a resin-treated, glass-insulated hub, to increase the effective collection area of *electrical fluid* while maintaining high insulation. A bent lever and chain connected this hub to a rotary, telescopic arm that could sweep a full circle and be extended or retracted to adjust the treated footprint. As in Configuration I, a terminal crown of points was expected to diffuse charge gently onto the vegetation.

In both configurations, the apparatus was entirely passive: it neither measured voltage nor current and it injected no external power. Since modern concepts such as electrical potential, current and standardized units were not yet available, Bertholon reasoned qualitatively with notions of "more" or "less" electricity. The individual components of each configuration are detailed in Table 1.

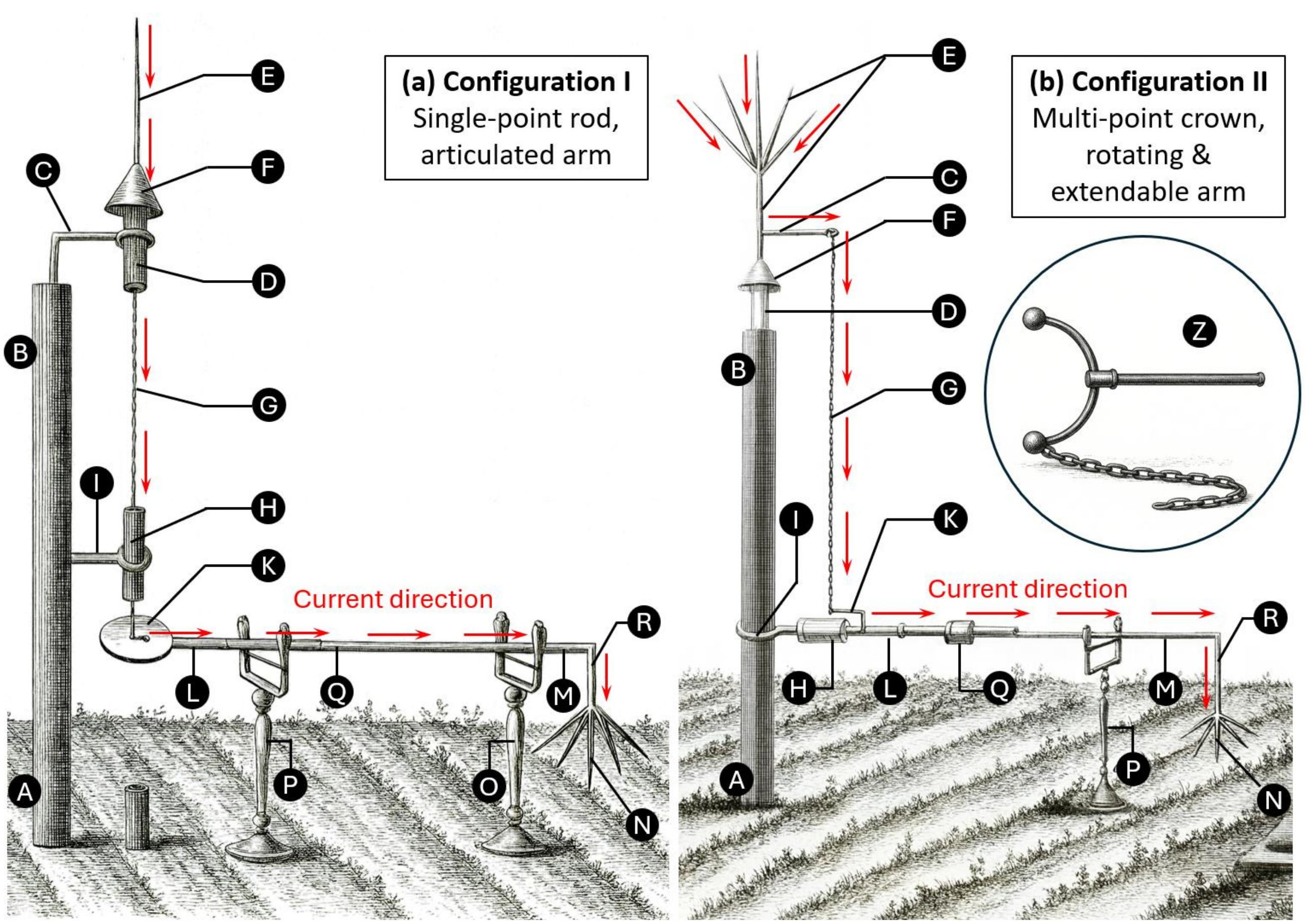

*Figure 1. Bertholon's electrovegetometer, a passive apparatus intended to channel "atmospheric electricity" toward crops. (a) Configuration I, featuring a single-point rod collector (E) at the mast head and an articulated arm that distributed charge through a multi-point crown (N). (b) Configuration II, equipped with a multi-point crown collector (E), a rotating and extendable arm and a larger treatment footprint terminating by a multi-point crown distributor (N). A neutralization device (Z), shown in the inset of (b), could provide safety and on/off control. Adapted from [Bertholon, 1783].*

*Table 1. Comparison between the design and operation of Bertholon's electrovegetometer (Figure 1): Configuration I ("single-point rod, articulated arm") and Configuration II ("multi-point crown, rotating and extendable arm"). The letters (A to Z) correspond to the captions of the original figures. Both devices were passive collectors-distributors (no field/charge/current measurement).*

| | Configuration I<br>Single-point rod, articulated arm | Configuration II<br>Multi-point crown, rotating & extendable arm |
|---|---|---|
| **Purpose** | Passive capture → Insulated conduction → Gentle diffusion over crops (no measurement). | |
| **Mast / Base** | Wooden mast (A-B): buried section fire-dried, tarred, wrapped in charcoal dust/cement, then masonry footing; above-ground part painted/bitumen-coated for weather resistance. | Mast (A-B) with resin-saturated head cylinder (C) treated with tar/pitch/turpentine to improve insulation from ground. |
| **Top insulation** | - Thick glass tube packed with bituminous mastic (D)<br>- Second glass insulator downstream (H). | Insulated rotary interface via resin insert (D) and glass/mastic sleeve (H). |
| **Collector (at mast head)** | Single-point rod (E) mounted in D; protected by a tinplate funnel (F). | Multipoint crown (E) to maximize charge capture. |
| **Conduction path (vertical)** | Conducting chain (G) suspended from E, guided through H to prevent leakage. | Bent iron lever (C) carrying chain (G) to the arm interface. |
| **Regulation / buffer** | Chain terminates at iron disk (K) acting as a small "condenser/regulator" to buffer atmospheric variations. | No separate disk; regulation implicit (focus on mobility and reach). |
| **Horizontal arm / distribution** | Articulated conductor arm (K-L-M-N) hinged at L and Q, supported on insulating trestles (O, P) with stretched silk cords. | Telescoping conductor arm (L-M) with its telescoping conductor (Q), supported on insulating trestle (P) with stretched silk cords. |
| **Terminal diffuser** | Crown of sharp metal points at the arm tip (N, R), releasing charge as non-disruptive corona over the crop canopy/soil. | |
| **Mobility / coverage** | Arm swings on hinges (L, Q) to cover a sector (row/strip). | Full 360° rotation via I-L-M & extend/retract (Q) to sweep a full circle and vary reach. |
| **Neutralization (on/off control)** | Apply a chain (or vertical rod) between K and the ground to break insulation and bleed charge. Also used to null effect when "excess" of electricity is presumed. | |
| **Operation (per Bertholon)** | Use when the air was "poor in electric fluid"; distributed via N; grounded at K when "excess" was observed. | |
| **Safety** | Sharp points released corona glows (no sparks). "C-shaped exciter" (copper/iron frame with glass handle + trailing ground chain) recommended to shunt charge during handling. | |

## I.3. From Enlightenment ideas to modern atmospheric electrodynamics

By the late 18th century, several essential aspects of atmospheric electricity were already recognised. Experiments by Stephen Gray and Charles François de Cisternay Du Fay had already distinguished conductors from insulators and introduced the idea of two opposite electrical states, later renamed "positive" and "negative" [Bailey, 2001]. Benjamin Franklin synthesised these ideas, defined the polarity convention and argued, on the basis of laboratory similarities and field experiments, that lightning was an electrical discharge of the same nature as sparks from machines [Williams, 2009]. His proposals were confirmed in France by Dalibard and Le Monnier who utilized elevated and well-insulated iron conductors to draw sparks during thunderstorms, hence confirming that storm clouds carry electrical charge [Winkler et al., 1898]. Soon afterwards, systematic electroscope measurements by John Canton in London and Giambattista Beccaria in Turin showed that even cloudless air was rarely neutral: slow "silent" discharges, diurnal variations of the field and luminous glows such as sailors' St Elmo's fire revealed that the lower atmosphere was persistently electrified [Herbert, 2012]. From a modern standpoint, St Elmo's fire was understood as a low-current corona discharge from sharp conductors in a strong ambient field [McGinness, 2025]. Yet these 18th-century insights were local and qualitative. A quantitative, global framework, in which thunderstorms, fair-weather regions and the ionosphere are coupled through a single atmospheric electric circuit, emerged only in the early 20th century, particularly through C. T. R. Wilson's work and later syntheses of the "global atmospheric electrical circuit" [Aplin, 2008].

Modern atmospheric electrodynamics frames these phenomena in terms of the global electric circuit (GEC). The Earth-ionosphere system behaves as a leaky spherical capacitor: thunderstorms and electrified clouds maintain the ionosphere at roughly +250/+300 kV relative to the surface, while "fair-weather" regions close the circuit through the weak electrical conductivity of the air [Tinsley, 2007], [Mareev, 2014]. Under clear undisturbed conditions, the atmospheric column carries a small, nearly steady downward conduction current density, described locally by Ohm's law:

$$J_z(z) = \sigma(z)\, E_z(z) \qquad \{1\}$$

where σ(z) is the electrical conductivity and $E_z(z)$ the vertical electric field (taken positive upward). Near the surface, typical values are $E_{0,atm} \approx -100$ to $-150$ V.m$^{-1}$ and $J_{0,atm} \approx -1$ to $-3$ pA.m$^{-2}$ [Guha, 2010]. Conductivity increases by many orders of magnitude with altitude (from about $10^{-14}$ S.m$^{-1}$ near the ground to $10^{-7}$ S.m$^{-1}$ in the lower thermosphere), so that $|E_z|$ decreases correspondingly and the potential V(z) rises from 0 V at the ground toward the ionospheric level of a few ×$10^5$ V [Kudintseva, 2016].

Under disturbed or stormy conditions, this simple vertical structure is strongly modified. Charge separation in thunderclouds produces large quasi-static fields at the surface (up to several kV.m$^{-1}$), enhances ion production and alters the conductivity profile [Kundt, 1999]. The vertical current density can increase into the nA.m$^{-2}$ range and strong, locally variable fields develop near sharp objects, creating conditions favourable for corona discharges and leaders [Standler, 1979]. These disturbed states form the storm branch of the global circuit are crucial for interpreting how a protruding conductor, such as Bertholon's electrovegetometer, might experience both modest fair-weather forcing and episodic, much stronger storm-time forcing.

***Table 2. Idealized near-surface regimes used as atmospheric boundary conditions***

| Parameter | | Fair weather conditions | Stormy weather conditions |
|---|---|---|---|
| **Electric field direction** | | Downward (negative) | Upward (positive) |
| **Vertical electric field $E_{0,atm}$** | | $-100 \rightarrow -150\ V.m^{-1}$ | $+1 \rightarrow +10\ kV.m^{-1}$ |
| **Electric potential** | **V(z)** | Increases with altitude $\frac{dV}{dz} = -E_z > 0$ | Decreases with altitude $\frac{dV}{dz} = -E_z < 0$ |
| | **V(1m)** | $+100\ V \rightarrow +150\ V$ | $-1\ kV \rightarrow -10\ kV$ |
| | **V(10m)** | $+1000\ V \rightarrow +1500\ V$ | $-10\ kV \rightarrow -100\ kV$ |
| **Conductivity at surface σ(0)** | | (1-5) × 10$^{-14}$ S.m$^{-1}$ | (1-10) × 10$^{-12}$ S.m$^{-1}$ |
| **σ variation with height (0-20 m)** | | Increases gradually (≈10×) due to ionization by cosmic rays and radioactivity | Increases sharply (≈100×) near ground due to field-induced ionization and space charge |
| **Conduction current density J** | | ≈ -2 pA.m$^{-2}$ (downward, negative) | +0.1→ +10 nA.m$^{-2}$ (upward, positive) |
| **Surface potential reference V(0)** | | 0 V (by convention) | 0 V (by convention) |
| **Breakdown electric field ($E_{bd}$)** | | ≈ 3.0 MV.m$^{-1}$ | 2.5-2.8 MV.m$^{-1}$ |

Owing to its limited height (**Figure 1**), Bertholon's apparatus can only interact with the lowest few meters of the atmosphere, where the idealised one-dimensional GEC picture breaks down and local processes dominate. Topography, aerosol loading, vegetation, humidity, surface conductivity and proximity to storm charge all influence the local electric field. For a passive collector-distributor such as the electrovegetometer, this near-surface layer controls both the magnitude of intercepted current and the likelihood of initiating discharges at the tips. For clarity, two idealized regimes can be distinguished, as summarized in Table 2:

- Fair-weather ("sunny") conditions, representing undisturbed atmospheric states with weak fields and very small conduction currents;
- Stormy-weather conditions, representing periods beneath electrified clouds or near active storm cells, where fields and currents can increase by orders of magnitude.

## I.4. Knowledge gap and research questions

Historical accounts describe Bertholon's electrovegetometer as a purely passive apparatus: the ambient atmospheric conduction current was collected on a sharp upper point and redistributed through a conductor by a lower crown of points, without any external power source or means of regulating or measuring the electrical state. What remains uncertain is whether such a passive configuration could, even in principle, generate ion or charge fluxes large enough to have physical significance at canopy height. No previous study has embedded the electrovegetometer explicitly in the framework of the global atmospheric electric circuit, with realistic fair-weather and storm-time fields, currents and conductivity profiles, nor quantified how far from the tips its influence on the electrical microenvironment of crops can extend.

A second gap concerns the absence of a systematic distinction between fair-weather and storm-affected regimes. Modern atmospheric electricity recognises that typical fair-weather conditions (e.g. $E_{0,atm} \approx -120\ kV.m^{-1}$, $J_{0,atm} \approx -2\ pA.m^{-2}$) are associated with weak, quasi-steady conduction currents, whereas storm-time conditions can support much stronger and more structured electric fields. Yet, existing historiographical and experimental discussions of the electrovegetometer do not quantify how these regimes would differentially drive a floating metal structure, nor do they estimate pre-corona upper bounds on the fields and currents attainable at the upper collector and the lower crown. This makes it difficult to judge whether historical reports of luminous "aigrettes" correspond to fields that plausibly exceed diffuse-corona onset thresholds.

Finally, the electrodynamic consequences of geometric uncertainty remain unexplored. Historical engravings constrain the general architecture of the electrovegetometer but not the precise mast height, arm length, tip spacing or apex angles. It is therefore unknown how strongly the background column current can be concentrated into the canopy layer by a realistic crown, over what distance above the soil the enhanced electric field and current density persist (e.g. within the first 0.2-0.5 m) and how sensitive these quantities are to collector and mast geometry.

The present work addresses these gaps by developing a simplified two-dimensional conduction model of the air-device system, in which the electrovegetometer is treated as a floating metal structure polarised by the global electric circuit. Within this pre-corona, ohmic framework, we calculate the potential, electric field and current density around idealised reconstructions of Bertholon's device under both fair-weather and storm-like forcing, for several plausible combinations of mast height, arm length and tip spacing. The simulations quantify (i) the maximum local fields and enhancement factors at single-point and multi-point collectors, (ii) the concentration of background column current into the canopy layer and the resulting vertical and horizontal gradients of $|E|$ and $|J|$ and (iii) the atmospheric and geometric conditions under which the computed fields approach or exceed empirical diffuse-corona onset criteria. Because space-charge accumulation and ionisation feedback are not modelled, these results should be interpreted as controlled, pre-onset upper bounds rather than predictions of discharge behaviour.

These considerations lead to the following research questions:

- Under realistic fair-weather forcing, what potentials, electric-field magnitudes and current densities are established around the upper collector and lower crown and over what length scales do these perturbations decay above the canopy?
- Under enhanced background fields representing pre-storm or storm-affected conditions, can a purely passive electrovegetometer reach electric-field levels at its tips that approach diffuse-corona onset thresholds, making historical luminous "aigrettes" physically plausible?
- How do field enhancement, current concentration into the canopy layer and proximity to corona-onset thresholds depend on mast height, arm length, tip spacing and collector design within the range consistent with historical descriptions?

# II. Model

All numerical computations were performed using GNU Octave (version 10.3.0), employing custom scripts for the finite-difference discretization, relaxation scheme, and post-processing.

## II.1. Physical assumptions and scope

The electrostatic environment around Bertholon's electrovegetometer is modeled as a quasi-steady, ohmic atmosphere carrying the global weather conduction current. The key assumptions are as follows:
- The atmosphere behaves as a linear, resistive medium characterized by a prescribed conductivity σ(x, z), dominated by its altitude dependence σ(z).
- The vertical conduction current is approximately conserved ($\vec{\nabla}.\vec{J} = 0$) in the absence of sources or sinks. Space charge ρ and ionization feedback are not explicitly resolved; instead, their net effect on conductivity is encoded in σ(z).
- The metallic parts of the device are treated as perfect conductors, each connected metallic region being equipotential. The structure is electrically floating: no external voltage is imposed, and its potential adjusts self-consistently under the imposed global-circuit forcing.
- The ground is represented as a conducting, equipotential boundary. The crop canopy is not explicitly resolved (fields are evaluated in the air above a flat ground), so fine-scale field distortions by individual plants are neglected.
- The problem is solved in two-dimensional [Cartesian / axisymmetric] geometry, assuming invariance in the out-of-plane direction.
- The wooden supports are represented as weak conductors (very low $\sigma_{wood}$) that can leak current but are highly resistiv²e compared to metal.

## II.2. Governing equations and conductivity profile

The electric potential $V(x,z)$ satisfies the stationary conduction equation {2} throughout all materials, while the electric field and current density follow from equations {3} and {4}. The atmosphere is described using the altitude-dependent conductivity profile of equation {5}, where $\sigma_0 = \frac{J_{0,\text{atm}}}{E_{0,\text{atm}}}$ is determined by the imposed near-surface atmospheric electric field $E_{0,\text{atm}}$ and the corresponding column current density $J_{0,\text{atm}}$. **Figure 2a** displays the resulting isopotential structure for values of $E_{0,\text{atm}}$ ranging from $-0.4\,kV.m^{-1}$to $10\,kV.m^{-1}$. Two physically distinct atmospheric regimes appear in this range and are therefore considered in the simulations:
- Fair-weather conditions, characterized by $E_{0,\text{atm}} \approx -120V.m^{-1}$ and $J_{0,\text{atm}} \approx -2\,pA.m^{-2}$, yielding weak potential gradients and gently curved isopotentials (left side of **Figure 2a**).
- Stormy-weather conditions, associated with much stronger positive electric fields, here represented by $E_{0,\text{atm}} = +5kV.m^{-1}$ and $J_{0,\text{atm}} = 5nA.m^{-2}$, which produce steep potential drops with altitude and tightly packed isopotentials (right side of **Figure 2a**).
- A conductivity scale height of $H_\sigma = 6km$ is adopted. In a vertically uniform atmospheric column, current conservation {6} then leads to the electric-field profile of equation {7}. Through Gauss's law, the corresponding diagnostic space-charge density is given by equation {8}.

$$\vec{\nabla}\cdot\left[\sigma(x,z)\,\vec{\nabla}V(x,z)\right] = 0 \qquad \{2\}$$

$$\vec{E} = -\vec{\nabla}V \qquad \{3\}$$

$$\vec{J} = \sigma\vec{E} \qquad \{4\}$$

$$\sigma_{\text{air}}(z) = \sigma_0 \exp\left(\frac{z}{H_\sigma}\right) \qquad \{5\}$$

$$J_z(z) = \sigma(z).E_z(z) = J_{0,atm} \qquad \{6\}$$

$$E_z(z) = -E_{0,atm}\exp\left(-\frac{z}{H_\sigma}\right) \qquad \{7\}$$

$$\rho(z) = \varepsilon_0\frac{dE_z}{dz} = -\varepsilon_0\frac{E_z(z)}{H_\sigma} \qquad \{8\}$$

The space charge density $\rho(z)$ is not explicitly solved. Instead, σ(z) is prescribed and the solver remains purely ohmic and pre-corona. For storm-like surrogate cases, the same form is considered although the magnitudes of $|E_{0,atm}|$, $|J_{0,atm}|$ and $\sigma_0$ are increased to match the idealized values in the "storm-like" regime. The solid regions are air with $\sigma_{\text{air}}(z)$ given by equation {5}, wood supports with $\sigma_{\text{wood}} \approx 10^{-15}\,S.m^{-1}$(strongly insulating but not perfectly) and metal conductor: $\sigma_{\text{metal}} = 1\times10^7 S.m^{-1}$. Owing to this strong conductivity mismatch, the metallic structure is treated as a single floating conductor with an effectively uniform internal potential.

## II.3. Geometry & Computational domain

The problem is solved in two dimensions over $x \in [-0.4,\,2.0]\,m$ and $z \in [0.0,\,4.0]\,m$ on a uniform grid of $N_x = 6000$, $N_z = 10\,000$ nodes (Δx = Δz = 400 µm). This 2D formulation represents an infinitely long structure in the out-of-plane direction; results are therefore cross-sections with unit depth.

The Bertholon-type apparatus is modeled as shown in **Figure 2b**. A wooden vertical mast, 2.0 m high and 0.15 m wide, forms the central support. A short horizontal wooden element is placed at $z = 1.5$ m and two wooden feet, each 0.5 m high and 1.5 cm thick, provide support at ground level beneath the conductor. The metallic conductor is represented by an L-shaped structure with triangular tips. At the top, a 0.30-m-high, 0.02-m-wide triangular tip located at $z = 2.2$ m serves as the upper single-point collector, with an apex angle of 4°. From the base of this tip, a vertical metallic segment extends 1.70 m downward with a thickness of 0.02 m. This segment connects to a horizontal arm 1.50 m long positioned approximately 0.5 m above the ground. At the end of this arm, a terminal crown composed of five triangular points (each 0.30 m high and 0.02 m at the base) is arranged with angular orientations of −40°, −20°, 0°, +20° and +40° relative to the vertical. This stylized geometry corresponds to sharp apex radii of order 1 mm.

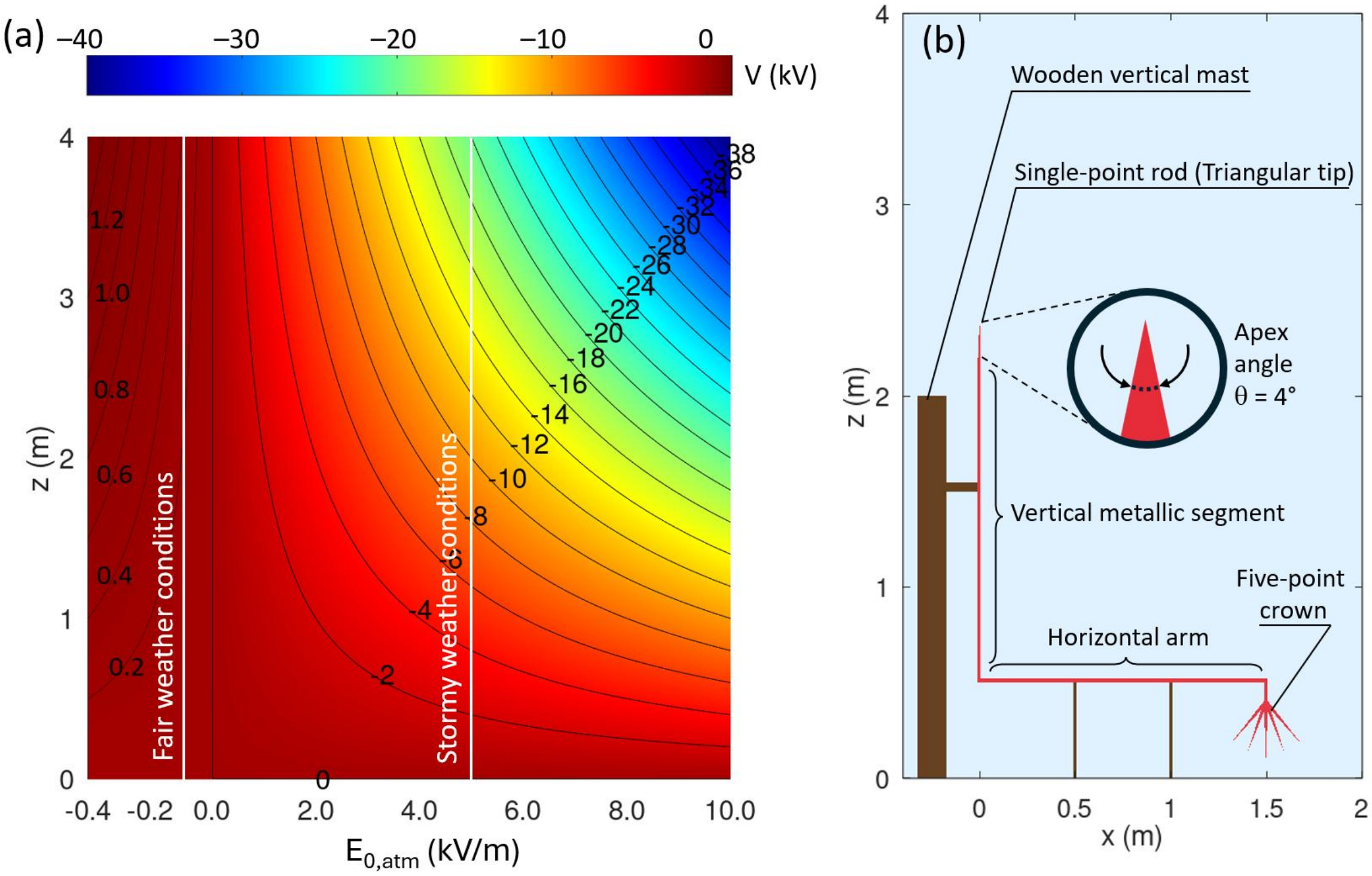

***Figure 2. (a) Electric potential $V(E_z, z)$ in the atmosphere above the ground, shown as colored contours (in kV) and black isopotential lines. The horizontal axis represents the imposed vertical electric field $E_z$, ranging from fair-weather values ($E_z < 0$) to storm-time values ($E_z > 0$). The two white vertical lines correspond to the fair-weather and stormy-weather atmospheric conditions for this study. (b) Geometrical model of Bertholon's electrovegetometer as implemented in configuration I, which includes the wooden insulating mast and supports (brown), the conductive L-shaped structure (red) and the multi-point crown. The computational domain represents a 2 m × 4 m vertical cross-section of the surrounding atmosphere.***

## II.4. Boundary conditions and numerical implementation

The top and bottom boundaries enforce the imposed vertical current density $J_{0,atm}$ following equations {9} and {10} while lateral boundaries are symmetric, as expressed by equation {11}.

$$\left.\frac{\partial V}{\partial z}\right|_{z=0} = -\frac{J_{0,atm}}{\sigma(0)} \qquad \{9\}$$

$$\left.\frac{\partial V}{\partial z}\right|_{z=H} = -\frac{J_{0,atm}}{\sigma(H)} \qquad \{10\}$$

$$\left.\frac{\partial V}{\partial x}\right|_{\text{left/right}} = 0 \qquad \{11\}$$

One air node is fixed at $V = 0$ to set the reference potential. This choice enforces the column current but does not impose a perfectly equipotential Earth at z = 0; near-ground potentials can therefore vary horizontally, an approximation acceptable here but not exact.

The governing equation {2} is discretized on a two-dimensional five-point finite-difference stencil. To ensure continuity of the current density across sharp changes in conductivity, the conductivity at each cell face is computed with equation {12} which provides the harmonic mean between adjacent cells. This averaging scheme preserves flux conservation and numerical stability in regions where materials of very different conductivities meet (e.g., air/wood or air/metal interfaces). The resulting discrete system is solved by Jacobi relaxation, in which the potential at node $(i, j)$ is updated iteratively according to equation {13} with boundary updates and gauge enforcement after each sweep.

$$\sigma_{\text{face}} = \frac{2\sigma_1\sigma_2}{\sigma_1 + \sigma_2 + \varepsilon} \qquad \{12\}$$

$$V_{i,j}^{(n+1)} = \frac{\sigma_e V_{i,j+1} + \sigma_w V_{i,j-1} + \sigma_n V_{i+1,j} + \sigma_s V_{i-1,j}}{\sigma_e + \sigma_w + \sigma_n + \sigma_s} \qquad \{13\}$$

Initialization uses the analytic fair-weather profile in equation {14} and metal regions are seeded with the average potential of neighbouring air cells. Iterations stop once the equation {15} is satisfied or when 2000 iterations are reached. This criterion yields good agreement with the 1D analytic column and current conservation (top vs bottom flux differ by <0.01%), but peak fields remain somewhat sensitive to grid resolution and tip geometry. They should be interpreted as approximate pre-corona values rather than fully converged quantities. Iterations continue until the maximum change In V between two successive steps falls below a prescribed tolerance, ensuring convergence of the electrostatic potential field.

$$V^{(0)}(z) = E_{0,\mathrm{atm}} H_\sigma \left(1 - e^{-\frac{z}{H_\sigma}}\right) \qquad \{14\}$$

$$\mathrm{Max}\left|V^{(n+1)} - V^{(n)}\right| < 10^{-3} \qquad \{15\}$$

Electric field and current density are computed after convergence using centered finite differences. The horizontal and vertical components of the electric field follow directly from the potential gradients (Equations {16}-{17}) and the field magnitude is obtained from equation {18}. The current-density components are then derived from Ohm's law (Equations {19}-{20}) using the local conductivity and the total current-density magnitude is computed from Equation {21}. These quantities are visualized in units of $kV.m^{-1}$ for the electric field and $pA.m^{-2}$ or $nA.m^{-2}$ for the current density, with solid regions masked.

$$E_x = -\frac{\partial V}{\partial x} \qquad \{16\}$$

$$E_z = -\frac{\partial V}{\partial z} \qquad \{17\}$$

$$|E| = \sqrt{E_x^2 + E_z^2} \qquad \{18\}$$

$$J_x = -\sigma . \frac{\partial V}{\partial x} \qquad \{19\}$$

$$J_z = -\sigma . \frac{\partial V}{\partial z} \qquad \{20\}$$

$$|J| = \sqrt{J_x^2 + J_z^2} \qquad \{21\}$$

# III. Results

## III.1. Fair-weather ("sunny") conditions

Here, we examine Configuration I of the electrovegetometer (**Figure 1a**), which consists of a conductive L-shaped structure combining a single-point collector at the upper extremity with a five-point crown near ground level. The response of this geometry is evaluated under fair-weather forcing, characterized by an imposed downward conduction current $J_{0,atm} \approx -2\ pA.m^{-2}$ and a background electric field $E_{0,atm} \approx -120V.m^{-1}$ applied at the lower boundary.

The steady-state potential, electric-field magnitude and current-density magnitude in the vicinity of the upper single-point collector are shown in **Figure 3**. **Figure 3a** illustrates that the atmospheric potential ranges from approximately 260 to 305 V over the vertical extent of the domain. Near the metallic point, the potential drops sharply to ≈260-265 V, whereas the surrounding air at the same altitude lies around 290-300 V. This 30-40 V lateral depression is tightly confined to a region only a few mm wide, forming a narrow potential funnel aligned with the tapered geometry of the rod. Because the conductor is electrically floating, its potential (≈240 V at the base of the point) emerges self-consistently as the value required to satisfy the imposed column current. Outside the immediate tip region, equipotential contours relax back to their nearly uniform fair-weather spacing.

**Figure 3b** reveals the resulting electric-field enhancement. The gradient associated with the 30-40 V drop over sub-cm distances produces local fields exceeding $20\ kV.m^{-1}$ with the simulation giving a peak value of $E_{\max} = 20.9\ kV.m^{-1}$. This corresponds to an enhancement factor of roughly:

$$\frac{E_{\max}}{|E_{0,atm}|} \approx \frac{2.1 \times 10^4}{1.2 \times 10^2} \sim 175$$

This value is purely due to geometric concentration at the apex. The elevated field decays rapidly: within 1 cm of the tip it falls to the kV/m range and within a few cm it returns to ambient fair-weather values.

**Figure 3c** shows the associated current-density magnitude. The strongest currents are confined to a narrow axial plume emerging from the single-point rod, with a peak of $J_{\max} = 1.36\ nA.m^{-2}$. This value is consistent with Ohm's law $J = \sigma E$, using the local fair-weather surface conductivity $\sigma_0 \approx 1.7 \times 10^{-14}\ S.m^{-1}$. Although small in absolute terms, such a current corresponds to a microscopic flux of:

$$\dot{N} \approx \frac{J_{\max}}{e} \approx 8 \times 10^9\ \text{charges.s}^{-1}.\text{pm}^{-2},$$

showing that even nanoampere currents involve billions of charge carriers per second. As expected in the ohmic, pre-corona regime, the spatial pattern of $|J|$ mirrors that of $|E|$ and merges smoothly into the background atmospheric current $J_{0,atm} \approx -2\ pA.m^{-2}$ over cm scales.

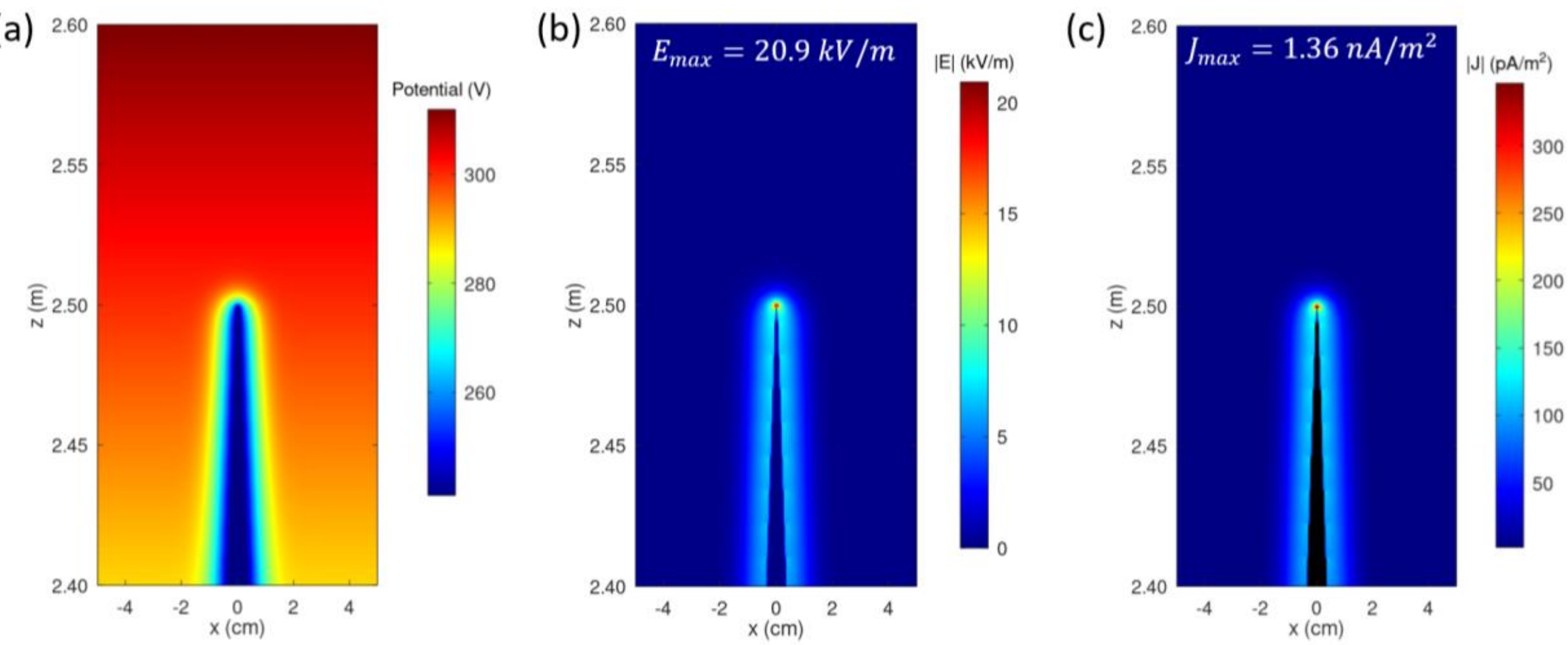

***Figure 3. Pre-corona response of the upper single-point rod collector (apex angle $\theta$ = 4°) under fair-weather forcing ($E_{0,atm} \approx -120V.m^{-1}$, $J_{0,atm} \approx -2pA.m^{-2}$). (a) Steady potential V showing strong equipotential compression near the floating metal apex. (b) Electric-field magnitude $|E|$with geometric enhancement to $E_{max} = 20.9\ kV.m^{-1}$. (c) Current-density magnitude $|J|$ peaking at $J_{max} = 1.36\ nA.m^{-2}$ and relaxing to the background within centimeters.***

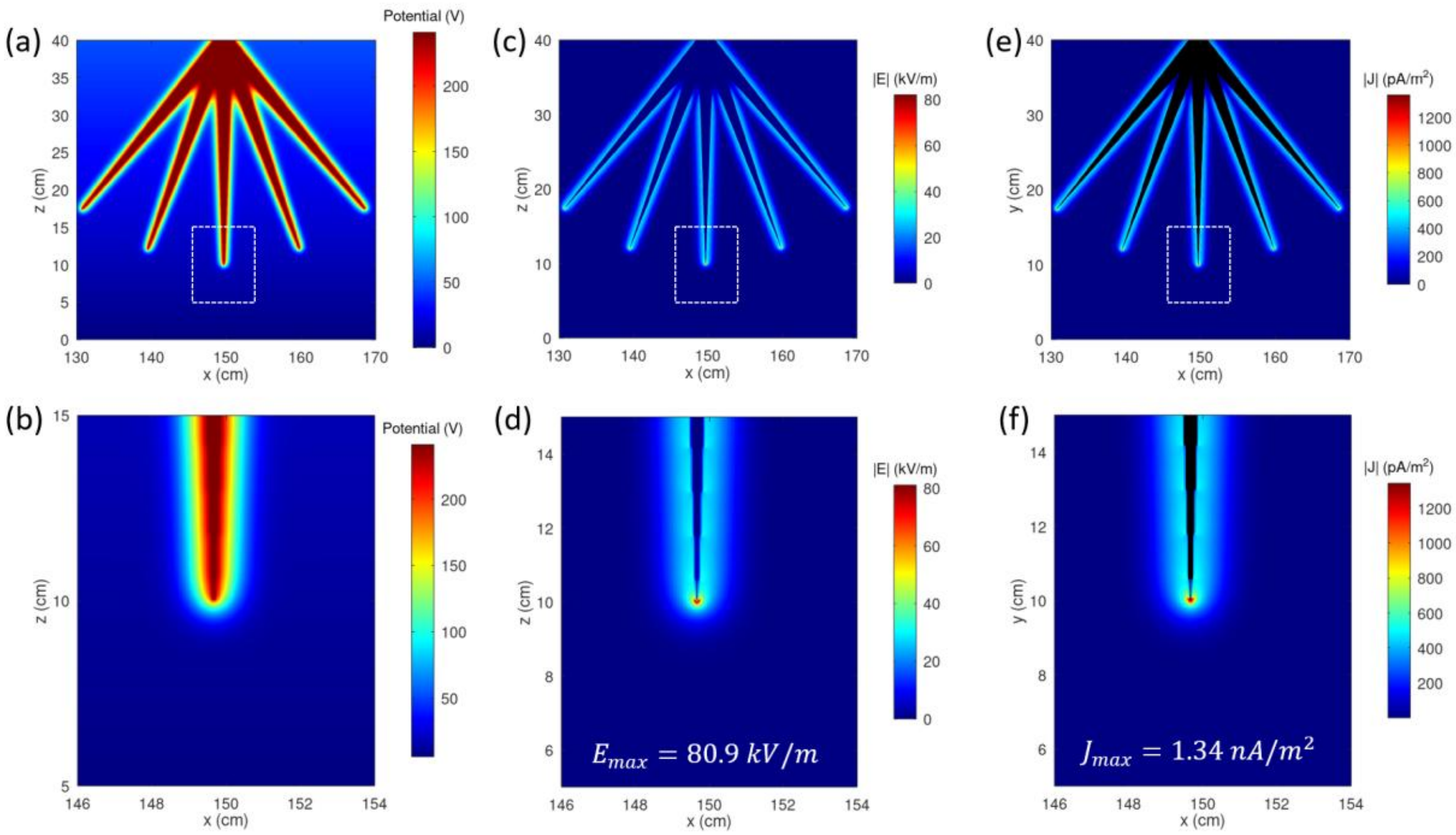

***Figure 4. Pre-corona response of the lower five-point crown under fair-weather forcing ($E_{0,atm} \approx -120V.m^{-1}$, $J_{0,atm} \approx -2pA.m^{-2}$) with apex angle $\theta$ = 4°. (a,b) Potential V showing local depressions around each apex and a narrow, deep potential funnel at the central tip. (c,d) Electric-field magnitude $|E|$ forming five interacting high-field lobes, with a localized maximum $E_{max} = 80.9\ kV.m^{-1}$. (e,f) Current-density magnitude $|J|$ revealing five focused conduction channels and a peak $J_{max} = 1.34\ nA.m^{-2}$. All values remain within the linear, conduction-dominated regime.***

The steady-state potential, electric-field magnitude and current-density magnitude surrounding the lower multi-point crown are represented in **Figure 4**. At the scale of the entire crown (**Figure 4a**), each of the five metallic points produces a distinct depression in the potential field. These local "funnels," typically 20-40 V deeper than the surrounding air at the same height, merge into a single broader perturbation a few cm away from the crown. The zoom near the central point (**Figure 4b**) shows a narrow, vertically elongated potential channel whose minimum lies around 250-260 V, corresponding to a lateral potential drop of 30-40 V over mm-to-cm scales. This results from both the sharp curvature of the central apex and the lateral confinement imposed by neighboring points of the crown.

The electric-field magnitude (**Figure 4c** and especially 4d) exhibits five distinct high-field lobes, one per apex. Their superposition forms a structured envelope rather than isolated maxima, reflecting strong inter-point interactions. Within the central zoom (**Figure 4d**) the field reaches $E_{max} \approx 80.9\ kV.m^{-1}$ , nearly 700 times the ambient fair-weather field. This field drops to a few kV/m within about 1 cm, returning to near-ambient values within several cm. The enhancement remains at least one order of magnitude below typical corona-onset thresholds for sharp conductors at ground pressure, confirming operation in the pre-corona, ohmic regime.

The current-density maps (**Figures 4e** and **4f**) indicate the structure of the electric field. Each point of the crown generates a narrow current plume that diffuses downward and merges with the background conduction current. The central apex again produces the highest value, $J_{\max} \approx 1.34\, nA.m^{-2}$ which corresponds to a microscopic flux similar to that estimated at the single-point rod collector, *i.e.* $\dot{N} \approx 8 \times 10^{9} charges.s^{-1}.m^{-2}$. Because the conductivity varies only weakly within the 5-15 cm region surrounding the crown, the structure of $|J|$ is governed almost entirely by geometric field shaping rather than atmospheric stratification.

Under fair-weather forcing, all simulated fields remain well below air-breakdown thresholds, confirming operation in a linear, ohmic pre-corona regime. Field enhancement arises solely from geometric concentration at the metallic points, producing smooth, monotonic patterns of electric field and current density consistent with $J = \sigma E$. These results define a quantitative baseline for potentials, enhancement factors and conduction plumes, against which stronger forcing can be compared. We now examine how the geometry responds under idealized storm-like conditions, where ambient fields increase dramatically.

## III.2. Stormy weather conditions

The response of the upper single-point rod is studied under an idealized storm-like forcing, using the parameters of Table 2. The background electric field is increased to $E_{0,\text{atm}} \approx 5000\, V.m^{-1}$ and the imposed conduction current reverses direction, reaching $J_{0,atm} \approx +5\, nA.m^{-2}$. This configuration provides a linear surrogate for storm-enhanced fields without explicitly modelling cloud charge, space charge or the three-dimensional electrostatic structure of thunderclouds. As a consequence, the fields obtained near the points represent upper bounds on the pre-corona state; a real atmosphere would transition to space-charge-dominated corona before such fields are maintained.

**Figure 5** shows the steady potential, electric-field magnitude and current-density magnitude around the upper tip for this storm-like scenario. The potential well around the apex (**Figure 5a**) is significantly deeper and more compressed than under fair weather, reflecting both the stronger forcing and the higher upward current. Local potential near the single point's apex fall below $-12\, kV$, compared with the $-260\, V$ range seen previously, indicating an order-of-magnitude intensification of the gradients. The resulting electric field (**Figure 5b**) reaches a peak of $E_{max} \approx 735\, kV.m^{-1}$ localized to a sub-cm region. This value is still below canonical breakdown estimates for air at ground pressure ($\sim 2.5 - 2.8\, MV.m^{-1}$) but it falls within the lower range of empirical corona-onset measurements for sharp conductors under humid or pre-storm conditions. Within 1 cm of the apex, the field drops to a few hundred kV/m and returns to background values within a few cm, preserving the strongly localized enhancement pattern observed in fair weather.

The corresponding current-density magnitude (**Figure 5c**) peaks at $J_{max} \approx 5.4\, \mu A.m^{-2}$ consistent with the scaling $J = \sigma E$, given the elevated conductivity at the tip's altitude. This represents an increase of nearly three orders of magnitude relative to fair-weather values ($\sim 1.36\, nA.m^{-2}$), yet the spatial structure remains a narrow axial plume diffusing smoothly into the background upward current.

Despite the large absolute fields and currents, the computed solution retains the qualitative signature of a linear conduction regime: equipotential surfaces continue to converge monotonically, $|E|$ increases smoothly toward the apex and no field clamping or plateau forms. This behaviour indicates that, within the limitations of a space-charge-free model, the system remains mathematically ohmic even at intensities where a real atmosphere would have already initiated corona. The results should therefore be interpreted as a near-threshold pre-corona state, with physical discharge onset inferred from external criteria, not from deviations within the simulated fields.

The response of the five-point crown distributor under storm-like forcing is shown in **Figure 6**. Compared with the fair-weather case, the stronger background field and reversed conduction current produce a markedly deeper and more confined potential well around each apex. On the scale of the full crown (**Figure 6a**), equipotential surfaces collapse tightly around all five metallic points, with local potentials in the zoom region (**Figure 6b**) falling to around $-10\, kV$. As in the previous single-point rod, these minima form narrow, vertically elongated channels, but the lateral confinement imposed by neighbouring peaks sharpens the gradients even further.

The corresponding electric-field magnitude (**Figures 6c-6d**) displays a set of 5 high-field lobes. Their superposition produces a structured envelope of enhanced field, with the central peak reaching $E_{\max} \approx 2845\, kV.m^{-1}$. This value is roughly 40 times larger than the storm-like single-point rod maximum and lies squarely within the empirical breakdown range for humid air at ground level. It also exceeds typical diffuse-corona onset estimates for sharp conductors. Because the model is strictly linear and excludes space-charge or ionization processes, these simulated fields should be interpreted as upper bounds on the pre-discharge state; a real atmosphere would begin corona formation and field clamping before such values were sustained. Spatially, the enhancement remains confined to a sub-cm region around the apex, decaying to the 100-of-kV/m range within ~1 cm and to background levels within a few cm.

The current-density maps (**Figures 6e-6f**) reproduce this structure. A sharply focused axial jet emerges from each point, with the central one producing $J_{\max} \approx 5.4\, \mu A.m^{-2}$, over three orders of magnitude above its fair-weather counterpart and consistent with the elevated $E_{\max}$ through $J = \sigma E$. Although extremely large for a conduction-only model, the spatial pattern remains smooth and monotonic, without the field clamping or non-linear features that would accompany real corona onset.

Taken together, these results show that the multi-point crown acts as an exceptionally efficient field concentrator under storm-like forcing, much more so than the upper single-point collector. However, because the model neglects space charge, ionization and discharge physics, the predicted MV/m-scale peaks represent idealized upper limits rather than realistic operational values; in practice, corona, streamer activity and associated charge regulation would intervene well before the simulated fields could be reached.

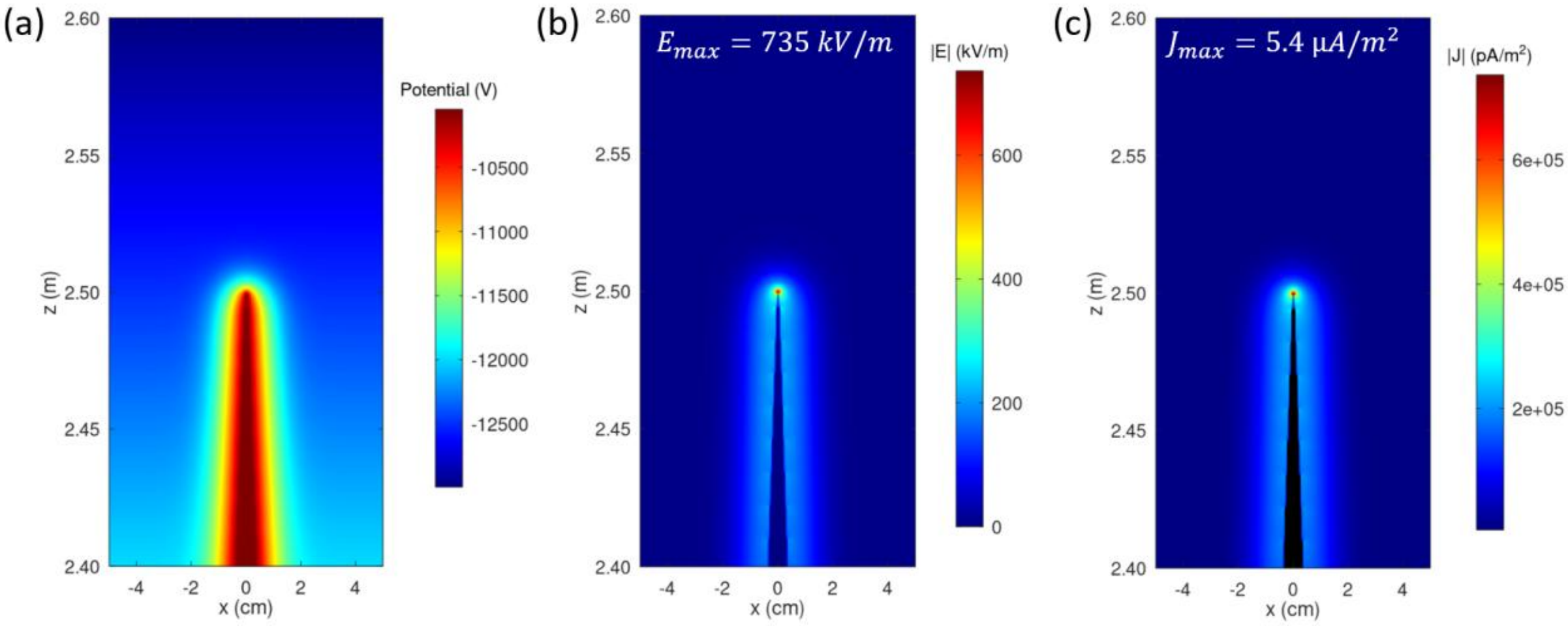

***Figure 5. Response of the single-point rod collector under idealized storm-like forcing ($E_{0,atm} \approx 5\ kV.m^{-1}$, $J_{0,atm} \approx 5\ nA.m^{-2}$). (a) Steady potential V showing deep compression of equipotentials near the floating metal apex. (b) Electric-field magnitude $|E|$ reaching $E_{max} = 735\ kV.m^{-1}$, approaching empirical corona-onset ranges. (c) Current-density magnitude $|J|$ peaking at $J_{max} = 5.4\ \mu A.m^{-2}$. Values represent pre-corona upper bounds within the linear, space-charge-free model.***

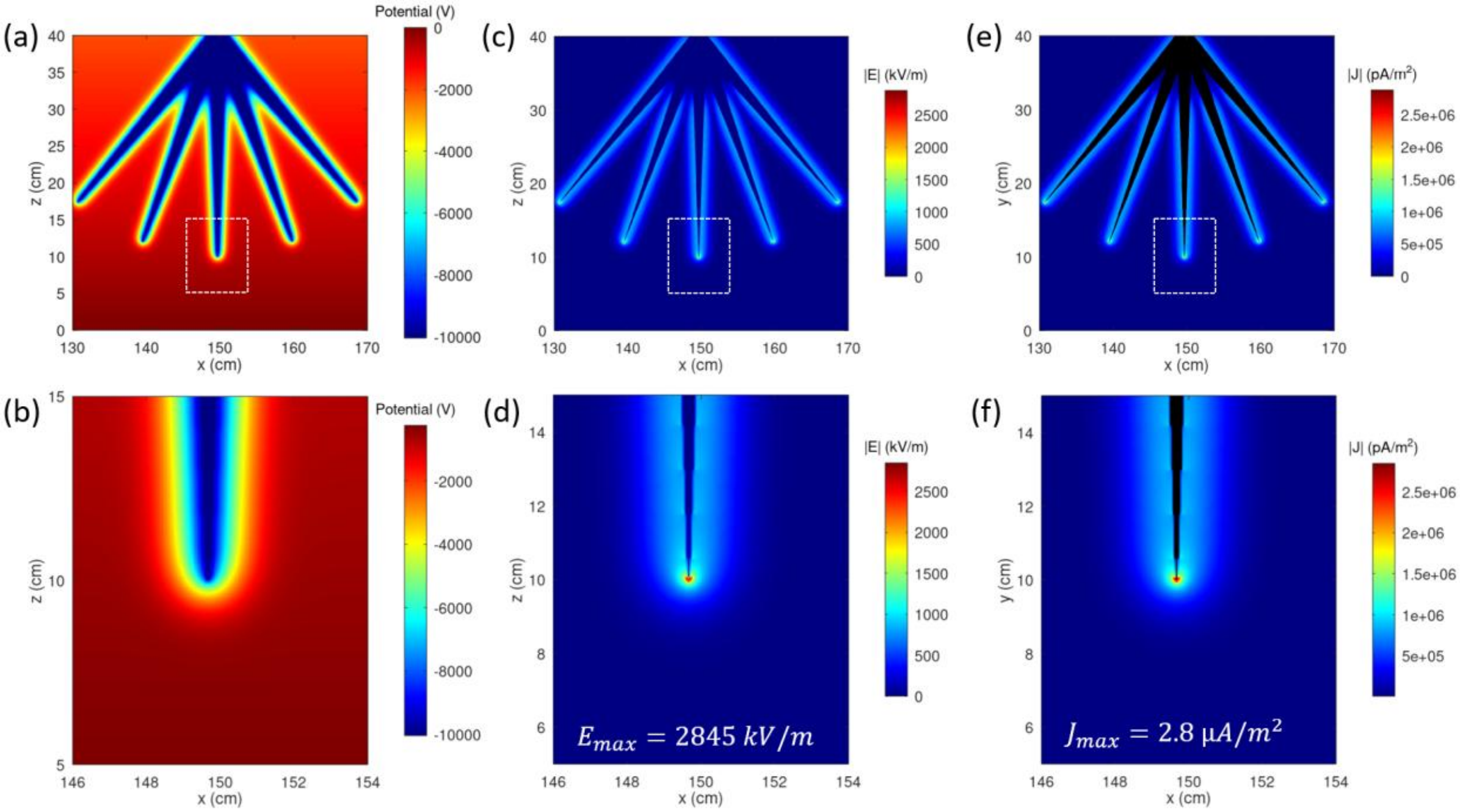

***Figure 6. Response of the lower five-point crown under idealized storm-like forcing ($E_{0,atm} \approx 5\ kV.m^{-1}$, $J_{0,atm} \approx 5\ nA.m^{-2}$) with apex angle $\theta$ = 4°. (a,b) Potential V showing a deeply collapsed well and strong confinement of equipotentials around each apex. (c,d) Electric-field magnitude $|E|$ with a localized maximum $E_{max} = 2.84\ MV.m^{-1}$, approaching empirical breakdown ranges. (e,f) Current-density magnitude $|J|$ peaking at $J_{max} = 2.8\ \mu A.m^{-2}$. These values represent upper-bound, pre-discharge fields within the linear, space-charge-free model.***

## III.3. Influence of apex angle

To evaluate the sensitivity of the electrovegetometer response to the sharpness of Bertholon's upper collector, a parametric sweep is achieved for different values of the apex angle θ of the single-point rod, keeping all other geometric and atmospheric parameters unchanged. The angle θ is defined here as the full opening angle of the triangular tip in the 2-D configuration (**Figure 2b**). Values from 2° to 90° (blunt, right-angle wedge) were explored. For each θ, the maximum electric-field magnitude ($E_{max}$) is extracted from the air adjacent to the upper tip and to the lower crown. Besides, the maximum conduction current density ($J_{max}$) is also extracted in those same regions. The results are summarized in **Figure 7**.

**Figure 7a** shows that the peak field near the single-point rod decreases monotonically as the tip is blunted. Under fair-weather forcing (red triangles), $E_{max}$ drops from roughly 20 kV/m for very sharp apices to 13 kV/m at 90°. The dependence is smooth and relatively weak: over the entire 0-90° range, the variation remains within a factor ≲1.5. The same trend holds under storm-like forcing (blue triangles), with $E_{max}$ decreasing from 800 kV/m for acute tips to 450 kV/m for the bluntest wedge. Thus, sharpening the apex enhances the local field, but only at the tens-of-percent level for realistic angles. This mild angular sensitivity is expected in a floating-conductor configuration dominated by large-scale column forcing: changing θ modifies the local curvature, hence the geometric amplification factor, but the overall potential of the metal adjusts self-consistently to satisfy the imposed atmospheric conduction current. In other words, the tip sharpness controls how a nearly fixed potential drop is concentrated into mm-cm gradients, rather than changing the global voltage available to the structure.

For comparison, **Figure 7a** also reports the maximum field at the five-point crown (stars). In both atmospheric regimes, the crown peak field remains essentially constant as θ varies. Under fair weather, it stays near 80 kV/m and under storm-like forcing near 2700-3000 kV/m. This insensitivity indicates that the lower crown is controlled almost entirely by its own apex geometry and by the ambient forcing, with only negligible coupling to the sharpness of the distant upper collector. Practically, this means that uncertainties in the historical manufacturing of the top point do not propagate into the near-canopy field enhancement produced by the terminal crown.

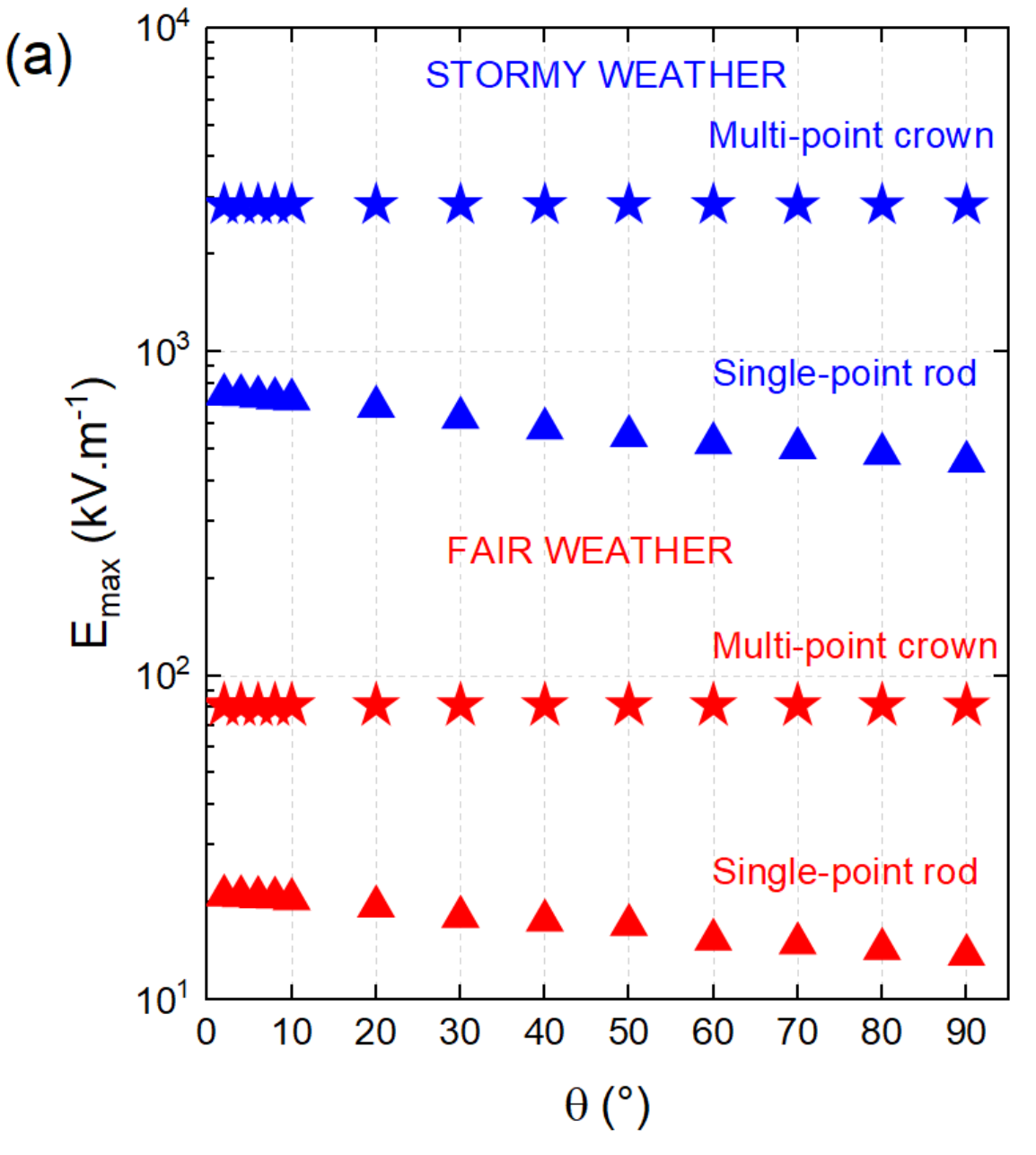

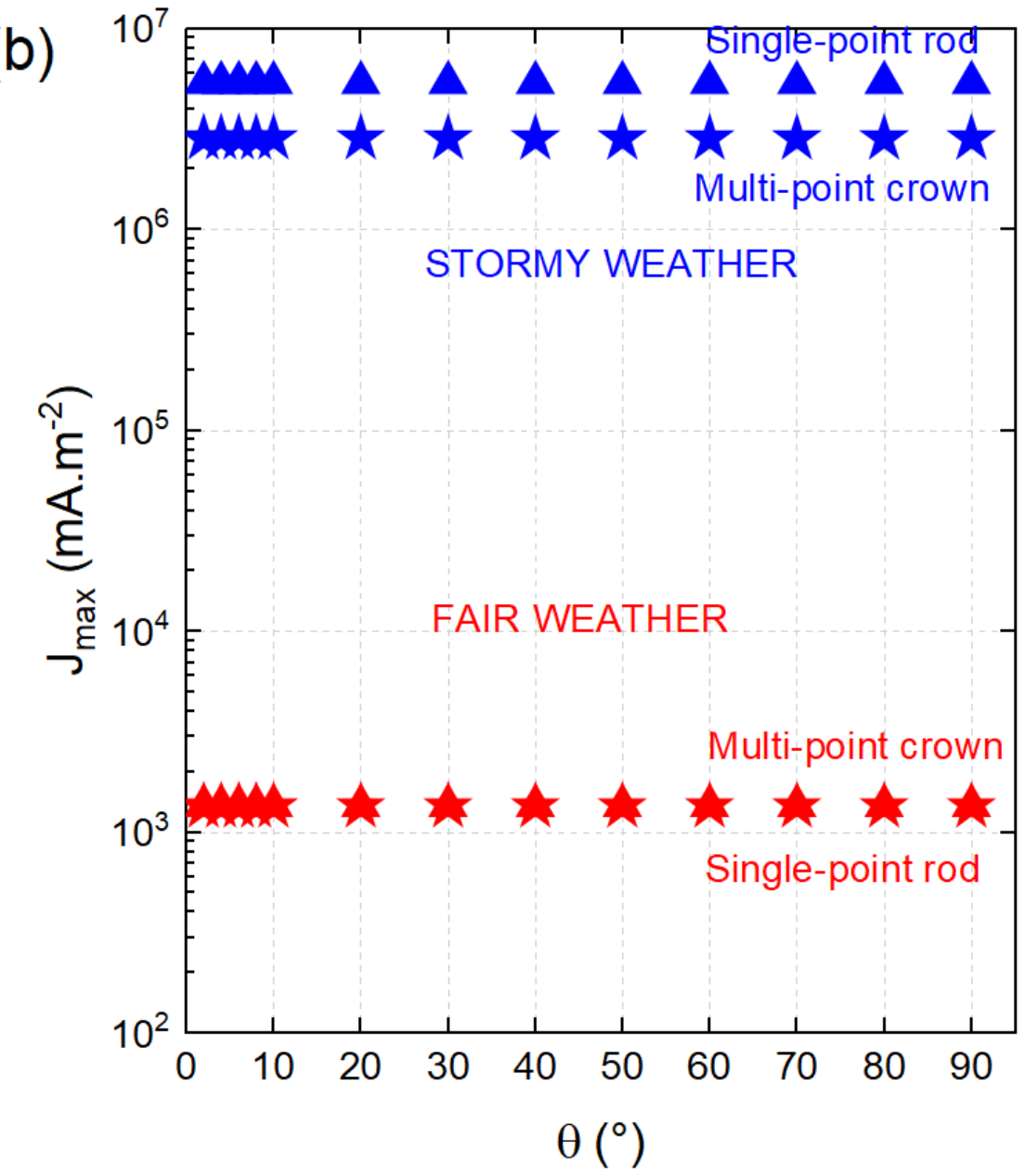

***Figure 7. (a) Maximum electric field $E_{max}$ and (b) current density $J_{max}$ as functions of apex angle for the upper collector, under fair-weather (red) and storm-like (blue) forcing. Fair-weather values show weak dependence on tip geometry, whereas storm-like conditions produce strong geometric amplification. The lower crown achieves the highest fields ($2.8\ MV.m^{-1}$) and currents ($5\ mA.m^{-2}$), representing upper-bound pre-discharge estimates in the linear ohmic model.***

The corresponding maxima in conduction current density are shown in **Figure 7b**. As predicted by the ohmic relation, $J_{max}$ near the upper tip follows the same monotonic decrease with θ as $E_{max}$. In fair weather (red triangles), the peak current density decreases gently (again by less than a factor of two) between sharp and blunt apices. In storm-like conditions (blue triangles) the same reduction occurs, but at values that are three orders of magnitude higher than in fair weather, reflecting the imposed increase in both background field and conductivity. As for the electric field, the crown current maxima (stars) are essentially flat with θ, confirming that the lower distributor behaviour is decoupled from the collector sharpness in this pre-corona regime.

An additional feature of **Figure 7b** is that, under storm-like forcing, the upper tip exhibits a larger $J_{max}$ than the crown even though the crown reaches higher $E_{max}$. This reversal arises from conductivity stratification: the upper collector operates at higher altitude where σ(z) is larger, so a given field produces a proportionally larger conduction current. In contrast, the crown's stronger field sits in a lower-σ layer and therefore carries a smaller ohmic current density despite its higher geometric amplification.

Altogether, the apex-angle sweep shows that the pre-corona predictions for the upper collector are robust to plausible uncertainties in tip sharpness. Whether the 18$^{th}$-century metal point was needle-sharp or moderately blunt changes $E_{max}$ and $J_{max}$ only by a modest factor, rather than by orders of magnitude. Therefore, the qualitative conclusions drawn from the baseline θ ≈ 4° geometry (namely that fair-weather operation remains safely sub-coronal while storm-time forcing can push the tip toward onset ranges) do not depend critically on an exact reconstruction of point shape. The strong enhancement produced by the multi-point crown, on the other hand, is governed by the crown's own sharpness and spacing and is effectively independent of the collector angle. Finally, it should be recalled that these angular trends are obtained in a 2-D wedge approximation. A fully 3-D conical tip would yield different absolute amplification factors, but the monotonic, weak sensitivity to θ is expected to persist, since it reflects the floating-potential adjustment and the localization of the potential drop rather than any peculiarity of the 2-D geometry.

## III.4. Influence of the conductor geometry

Configuration I of the electrovegetometer appears effective under storm-time conditions but essentially inactive in fair weather. To explore whether alternative geometries could improve its fair-weather response, four variants of the conductor layout were examined:

- The configuration I with single-point rod collector and crown distributor
- The configuration II: the single upper point is replaced by a small crown of sharp tips mounted at the mast head, while the lower distributor remains unchanged.
- The extended configuration I: it retains a single upper point but extends the metallic mast by approximately 1 m, so that the tip now protrudes higher into the atmospheric column. The lower crown and supports are unchanged. This configuration does not correspond to a specific historical drawing but serves to test the sensitivity of the device to collector height and to the associated changes in background potential and conductivity.
- The isolated crown configuration: the entire metallic mast-arm assembly is removed so as to leave only the five-point crown at canopy height. This "minimal" configuration provides a control case in which the crown interacts directly with the ambient field without any guidance or concentration by an elevated conductor.

For the four configurations, the maximum electric field values obtained at the lower crown are presented in **Figure 8a** under fair-weather and storm-like forcing. In fair weather, all configurations yield nearly identical peak fields of order $10^2$ kV.m$^{-1}$, indicating that the local crown geometry dominates field enhancement and that the presence, shape or height of the upper collector has only a minor influence in the weak-field regime. Under storm-like forcing, however, the three mast-bearing configurations (I, II and extended-I) all reach comparable peak fields of about $2.5\text{-}3 \times 10^3$ kV.m$^{-1}$, whereas the crown-only configuration remains significantly lower (roughly one half of that value). This contrast shows that the elevated metallic mast efficiently funnels the storm-enhanced column potential into the crown region, but that replacing a single-point collector by a multi-point collector or modestly extending its height does not materially change the local maximum field at the distributor.

The corresponding maxima in conduction current density, $J_{\max}$, are shown in **Figure 8c** on a logarithmic scale. In fair-weather conditions, all four configurations cluster around $J_{\max} \approx 10^{-3}$ $\mu$A.m$^{-2}$, confirming that the weak, sub-coronal currents are essentially set by the ambient atmospheric conduction current and by the crown geometry. When storm-like forcing is applied, $J_{\max}$increases by nearly three orders of magnitude, reaching a few $\mu$A.m$^{-2}$ for the three mast-bearing configurations, while the crown-only case again remains lower by about a factor of two. The parallel behaviour of $E_{\max}$ and $J_{\max}$ across the four configurations is consistent with the ohmic scaling $J = \sigma E$and confirms that the main role of the mast and upper collector is to increase the amount of background current intercepted and channelled through the crown, rather than to introduce strong qualitative differences between the specific collector designs.

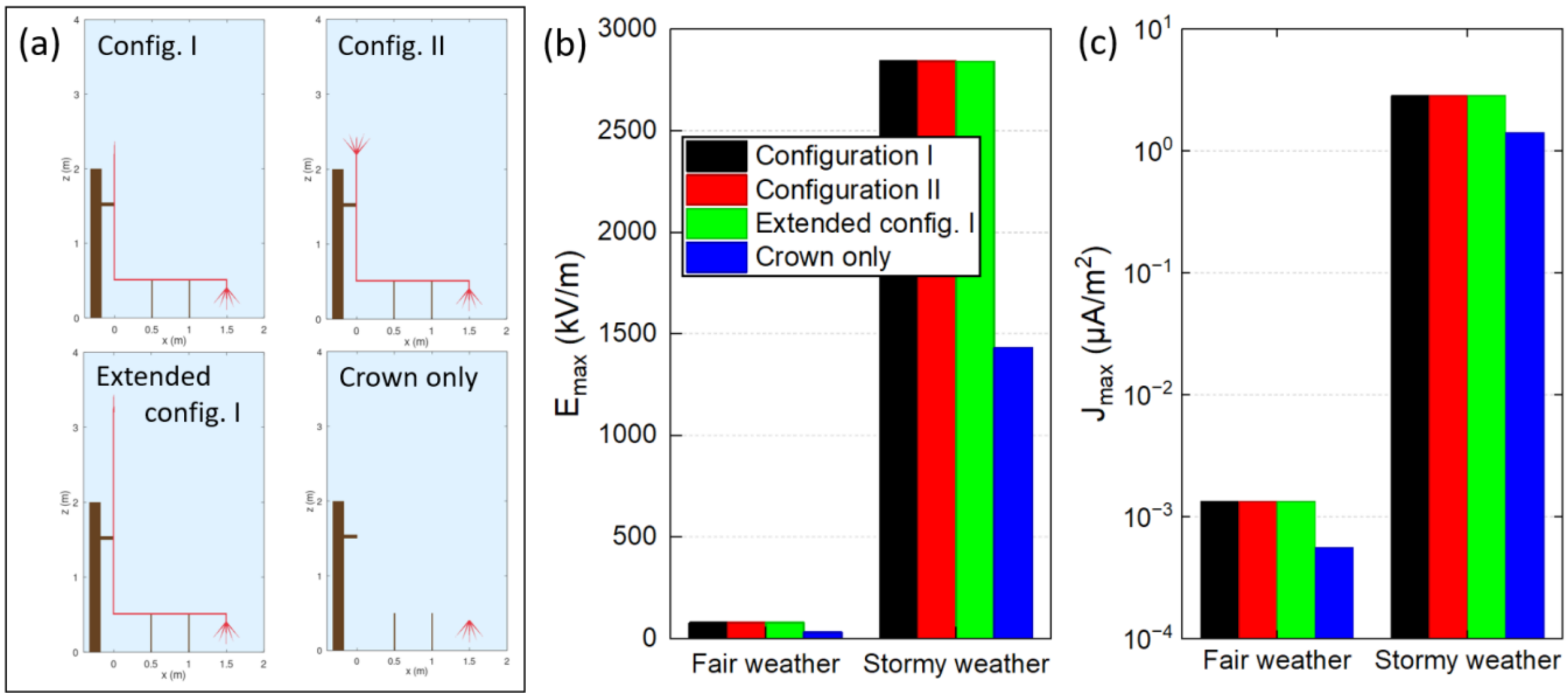

*Figure 8. (a) Four distinct geometries of the electrovegetometer used in the simulations. Brown regions denote wooden supports, red lines metallic conductors. (b) Maximum electric field at the lower crown distributor, $E_{max}$, under fair-weather and storm-like forcing, showing that all mast-bearing configurations produce similar peak fields while the crown-only case remains lower in the storm regime. (c) Corresponding maximum conduction current density, $J_{max}$, on a logarithmic scale, illustrating the three-order-of-magnitude increase from fair to storm conditions and the enhanced currents obtained when an elevated metallic mast is present.*

# IV. Discussion

The simulations presented above provide a first quantitative bridge between Bertholon's qualitative description of his electrovegetometer and modern concepts of the global atmospheric electric circuit and pre-corona field enhancement. Several points emerge regarding (i) the physical plausibility of Bertholon's original narrative, (ii) the expected magnitude and localisation of electrical effects near the canopy and (iii) the limitations of the present modelling framework and its implications for both historical interpretation and modern "electroculture" claims.

## IV.1. Fair-weather operation: localized but modest enhancement

Under fair-weather forcing, the reconstructed electrovegetometer behaves as a weak perturbation of the global conduction current. The upper collector and lower crown both generate strong local field enhancements (by two to three orders of magnitude relative to the ambient $\approx 10^2$ V.m$^{-1}$ field) but these enhanced regions remain confined to mm-cm neighbourhoods around the tips. The single upper point concentrates the background potential drop into a narrow funnel and attains peak fields of order a few $\times 10^4$ V.m$^{-1}$, while the five-point crown reaches about $8\times10^4$ V.m$^{-1}$ near the central apex.

From a purely electrodynamic standpoint, these fields are high enough to significantly distort ion trajectories and to locally increase ionisation and attachment rates, if the atmosphere is already weakly ionised. However, the associated ohmic current densities remain in the pA.m$^{-2}$ to low-nA.m$^{-2}$ range, only moderately above the background conduction current. The total current that can be channeled through a realistic crown footprint therefore remains extremely small.

For the plant microenvironment, this implies that, in undisturbed weather, the device could plausibly create narrow "hot spots" of enhanced field and ion flux immediately around each tip, but these perturbations would decay to near-background values over distances comparable to or smaller than, typical leaf and canopy length scales. Any direct electrodynamic forcing on plant tissue (through induced surface charges, field-driven ion flows at the leaf boundary layer or modification of aerosol deposition patterns) would thus be highly localised and likely intermittent.

Consequently, the fair-weather regime appears compatible with Bertholon's qualitative idea of a continuous but "gentle" influence of atmospheric electricity, yet the present results suggest that this influence is physically subtle: the electrovegetometer does not act as a strong injector of charge or as a significant sink/source in the global circuit. Rather, it reshapes the local pre-existing conduction current into narrow plumes without greatly increasing its magnitude.

## IV.2. Storm-time forcing and the plausibility of luminous "aigrettes"

Under idealised storm-like forcing, the same geometry operates in a very different regime. When the background field and conductivity are increased to values representative of pre-storm or storm-affected conditions, the upper point and lower crown can reach peak electric fields ranging from several $10^5$ V.m$^{-1}$ up to the MV.m$^{-1}$ scale in the linear model. These levels are comparable to or exceed, empirical corona-onset thresholds for sharp conductors at ground pressure and approach conventional breakdown estimates.

It is important to emphasise that, because the model neglects space charge and ionisation feedback, these $MV.m^{-1}$-scale values should be interpreted as *upper bounds* on the pre-corona state rather than as realistic sustained fields: in a real atmosphere, corona would be triggered earlier and the resulting space charge would partially screen the tips, limiting further field growth. Nonetheless, the fact that the ohmic pre-corona solution naturally approaches corona-onset thresholds under enhanced GEC forcing is significant. It provides a quantitative rationale for Bertholon's reports of faint luminous glows ("aigrettes lumineuses") at the points under disturbed weather and for historical observations of St Elmo's fire on ship masts and church spires.

In this picture, the electrovegetometer can act as a facilitator for corona discharges during periods when the atmospheric column is already highly stressed by nearby storm charge. The elevated mast ensures that the upper collector samples a region of slightly higher conductivity and potential, while the lower multi-point crown concentrates this potential drop at canopy height. The simulations show that storm-time currents may increase by roughly three orders of magnitude compared with fair weather while retaining smooth, monotonic conduction-like profiles around the tips. Once corona onset is reached, ion production and charge transport near the canopy could be substantially greater than in the fair-weather regime, potentially producing visible glows and enhanced ion fluxes.

However, this enhanced operation would remain episodic, tied to the presence of nearby electrified clouds and may not correspond to the benign, steady-state "improvement" of vegetation that Bertholon had in mind. Moreover, corona can produce localised chemical and mechanical stresses (ozone, NOx, UV, micro-scale heating and ion-wind shear) whose impacts on plants are complex and not necessarily beneficial. The present model cannot address these aspects and thus cannot directly support or refute any claimed agronomic benefits of storm-time operation.

## IV.3. Robustness to geometric uncertainties and the role of the mast

The parametric sweeps indicate that several aspects of the device response are relatively robust to historical uncertainties in geometry. Varying the apex angle of the upper collector over a wide range only alters the maximum field and current density near the tip by factors of order unity, not by orders of magnitude. This reflects the fact that the global potential available to the floating structure is set by the imposed atmospheric column, while the tip sharpness controls how that potential drop is localized on mm scales. Likewise, replacing the single upper point by a small crown or extending the mast modestly, barely affects the peak fields at the lower crown in either fair or storm-like conditions. The simulations show that, in the storm regime, all mast-bearing configurations (Configurations I, II and extended-I) yield similar $E_{max}$ and $J_{max}$ at the canopy-level crown, whereas an isolated crown with no mast reaches significantly lower maxima. This suggests that the primary electrodynamic role of the mast is to intercept an enhanced portion of the background column current and potential, funnelling it into the distributor, rather than to sensitively encode the details of the collector design.

From a historical reconstruction perspective, this is reassuring: even if Bertholon's actual hardware deviated from the stylised geometry used here, the qualitative conclusions about fair- versus storm-time operation, localisation of fields and the relative importance of collector versus distributor are unlikely to change drastically. The multi-point crown near the crop canopy emerges as the dominant feature for local field enhancement, while the mast and collector primarily condition how much "voltage" is delivered to it.

## IV.4. Limitations of the present model

Despite its usefulness as an exploratory tool, the present 2-D, linear, ohmic model has several important limitations that constrain the interpretation of the results:

- The model assumes a prescribed conductivity profile and linear Ohm's law, with no explicit treatment of charge production, attachment, recombination or drift beyond what is encoded in $\sigma(z)$. Once fields approach corona-onset thresholds, this approximation breaks down. The simulated peak fields in the storm-like regime should therefore be viewed as indicative of where onset is likely, not as literal predictions of sustained $MV.m^{-1}$ values.
- The cross-sectional configuration of the electrovegetometer represents an infinitely long structure, leading to wedge-like tips rather than realistic 3-D conical points. This affects local amplification factors and field line topology near the apices. A full 3-D simulation would likely yield different absolute $E_{max}$ values and possibly broader high-field regions, though the basic trends (localised enhancement, decimetric decay to background, relative insensitivity to collector apex angle) should remain qualitatively similar.
- The wooden supports are modelled with fixed, low conductivities and the soil is treated implicitly through boundary conditions rather than as a stratified, moisture-dependent conductor. In practice, however, Bertholon devoted considerable attention to improving insulation: the buried mast section was fire-dried, tarred, wrapped in charcoal dust and cement, then set in masonry, while the above-ground part was painted or bitumen-coated; the mast head was further protected by resin-saturated inserts and glass/mastic sleeves and the horizontal arms were carried on insulated trestles with silk cords (Table 1). These measures likely reduced leakage and increased support resistance, yet their properties still depended on humidity, ageing and contamination, so leakage paths and the floating potential (and thus $E_{max}$ and $J_{max}$) may have shifted only by moderate factors.
- The "fair-weather" and "storm-like" regimes are represented by simple, horizontally uniform profiles with prescribed $J_{0,atm}$ and $E_{0,atm}$. Actual atmospheric electric environments over agricultural fields are influenced by topography, nearby structures, aerosol loading and horizontal inhomogeneities in storm charge. The local field at the electrovegetometer could therefore deviate significantly from the idealised profiles adopted here, especially under complex convective conditions.

- The simulations do not include explicit representations of vegetation. Leaves, stems and canopy structure can distort the field and act as additional conductors or partial insulators, redistributing currents and modifying the near-surface potential landscape. They can also participate in charge exchange through surface conduction films, stomatal apertures or wetting layers. The present results therefore provide a first-order picture of the field “in the air”, not a complete description of plant-field coupling.

## IV.5. Implications for the history and modern practice of “electroculture”

Within these limitations, the modelling results support a nuanced re-assessment of Bertholon’s electrovegetometer. On the one hand, the device is physically capable of generating strong local field enhancements and, under storm-time conditions, of approaching or exceeding corona-onset thresholds, making luminous “aigrettes” plausible. On the other hand, in its primary fair-weather operating state (which would dominate day-to-day exposure) the structure appears to act mainly as a passive field concentrator with small total currents and highly localised perturbations.

This dual picture suggests that Bertholon’s conceptual intuition about the existence of a persistent electrical influence on vegetation was not entirely misplaced, but that his expectations regarding the strength and agronomic consequences of this influence might have been optimistic. The electrovegetometer is better interpreted as a delicate probe of the global electric circuit in the near-surface layer and as a demonstrator of field concentration and corona phenomena, than as a robust agricultural “enhancer” in the modern engineering sense.

For contemporary discussions of “electroculture”, a term that has recently resurfaced in popular and sometimes pseudoscientific contexts, the present work underscores the need for careful, quantitative modelling and controlled experiments. Devices that resemble historical electrovegetometers may indeed modify the local electric environment around plants, but the magnitude, sign and biological relevance of these modifications cannot be inferred from qualitative analogies alone. They require a combination of atmospheric-electric measurements, detailed plant-physiology studies and models that extend beyond the pre-corona, linear regime considered here.

# V. Conclusion

Re-examining Bertholon’s 18th-century electrovegetometer within the framework of modern atmospheric electrodynamics, this study has developed a quasi-steady, two-dimensional, resistive model of the air-device system, in which the atmosphere is treated as an ohmic medium carrying the global conduction current and the metallic structure is represented as a floating conductor supported by leaky wooden insulators. Solving the stationary conduction equation with altitude-dependent conductivity and realistic configurations has yielded quantitative estimates of potentials, electric fields and current densities around the electrovegetometer under idealised fair-weather and storm-like forcing.

The main findings can be summarised as follows:

- Under typical fair-weather conditions, the electrovegetometer’s sharp points enhance the local electric field by two to three orders of magnitude relative to the background atmospheric field, reaching tens of $kV.m^{-1}$ at the upper single point and about 80 $kV.m^{-1}$ at the lower multi-point crown. The associated current densities remain in the $pA\text{-}nA/m^2$ range, consistent with weak conduction in a slightly ionised medium. Field and current perturbations are strongly localised and decay to near-ambient values within cm.
- When the background field and conductivity are increased to values representative of pre-storm or storm-affected conditions, the same geometry yields peak fields in the $10^5$-$10^6$ $V.m^{-1}$ range. These levels approach or exceed empirical corona-onset thresholds and encroach on breakdown estimates, particularly at the lower crown. Within the limits of a linear, space-charge-free approximation, these results indicate that Bertholon’s device could have produced corona glows and enhanced ion fluxes during disturbed weather, providing a plausible physical basis for historical reports of luminous “aigrettes”.
- Parametric variations of the upper apex angle and collector geometry show that fair-weather and storm-time peak fields and currents near the lower crown are largely insensitive to the fine details of the collector, as long as an elevated mast is present. The mast’s principal function is to intercept a portion of the atmospheric column potential and funnel it to the distributor, while the multi-point crown near canopy height remains the dominant feature controlling local field enhancement.
- The results are contingent on the assumptions of a 2-D geometry, linear ohmic conduction, prescribed conductivity profiles and simplified material properties. They should therefore be viewed as pre-corona upper bounds rather than fully realistic predictions of discharge behaviour. Within these limits, the study suggests that Bertholon’s electrovegetometer was physically capable of concentrating atmospheric electricity and occasionally producing visible corona under storm-time forcing, while its fair-weather influence on plants was likely subtle, localised and difficult to quantify with $18^{th}$-century instruments.

Finally, this work illustrates how modern numerical modelling of the global atmospheric electric circuit, even in a simplified pre-corona form, can sharpen our understanding of historical electrotechnical devices and their plausible modes of operation. Future extensions could include fully 3-D simulations, explicit treatment of space charge and corona dynamics, more realistic representations of soil and vegetation and coupled experiments in controlled atmospheric-electric environments. Such studies would not only refine the retrospective assessment of Bertholon’s electrovegetometer but also inform contemporary debates on the physical and biological significance of weak atmospheric-electric perturbations in agricultural settings.